\documentstyle[12pt]{article}
         \makeatletter
         \def\thefigure{\@arabic\c@figure}\def\fps@figure{tbp}
         \def\ftype@figure{1}\def\ext@figure{lof}
         \def\fnum@figure{\protect\footnotesize Fig.\ \thefigure}
         \def\thetable{\@arabic\c@table}
         \def\fps@table{tbp}\def\ftype@table{2}\def\ext@table{lot}
         \def\fnum@table{\protect\footnotesize Table \thetable}
         \def\@listI{\leftmargin\leftmargini\parsep=0pt\itemsep=0pt}
         \def\thebibliography#1{\section{References}\vspace*{-10pt}\list
          {[\arabic{enumi}]}{\settowidth\labelwidth{[#1]}\leftmargin\labelwidth
          \advance\leftmargin\labelsep
          \usecounter{enumi}}
          \def\newblock{\hskip .11em plus .33em minus .07em}
          \sloppy\clubpenalty4000\widowpenalty4000
          \sfcode`\.=1000\relax}
         
         \def\@nomath#1{\ifmmode \fi}
         \def\mmycite{\@ifnextchar [{\@tempswatrue\@mmycitex}
             {\@tempswafalse\@mmycitex[]}}
         \def\@mmycitex[#1]#2{\if@filesw\immediate%
         \write\@auxout{\string\citation{#2}}\fi
           \def\@citea{}\@mmycite{\@for\@citeb:=#2\do
             {\@citea\def\@citea{,}\@ifundefined
                {b@\@citeb}{{\bf ?}\@warning
                {Citation `\@citeb' on page \thepage \space undefined}}%
         \hbox{\csname b@\@citeb\endcsname}}}{#1}}
         \def\@mmycite#1#2{{{\scriptsize#1}\if@tempswa , #2\fi}}
         \def\mycite#1{$^{\protect\mmycite{#1}}$}
         \def\@cite#1#2{{#1\if@tempswa , #2\fi}}
         \def\thesection {\arabic{section}}
         \def\section#1{\addtocounter{section}{1}\setcounter{subsection}{0}
              \bigskip\medskip{\noindent\bf\thesection.\ #1}
              \medskip}
         \def\thesubsection {\arabic{section}.\arabic{subsection}}
         \def\subsection#1{\addtocounter{subsection}{1}
              \medskip{\noindent\thesubsection.\ #1}
              \medskip}
         \makeatother
\begin{document}


\begin{center}
{\bf Liquid Drop Model and
Quantum Resistance against Noncompact Nuclear Geometries\\}
\bigskip
\bigskip
{J. T\~oke and W.U. Schr\"oder\\}
{\it Department of Chemistry\\ 
University of Rochester, Rochester, New York 14627}
\end{center}
\bigskip
\smallskip
{\footnotesize
\centerline{ABSTRACT}
\begin{quotation}
\vspace{-0.10in}
The importance of quantum effects for exotic nuclear shapes
is demonstrated.
Based on the example of a sheet of nuclear matter
of infinite lateral dimensions but finite thickness, 
it is shown that the quantization of states
in momentum space, resulting from the confinement
of the nucleonic motion in the conjugate geometrical space,
generates a strong resistance against such a confinement and
generates restoring forces driving the systems toward compact
geometries. In the liquid drop model, these quantum effects are
implicitly included in the surface energy term, via a choice
of interaction parameters, an approximation
that has been found valid for compact shapes, but has not yet
been scrutinized for exotic shapes.
\end{quotation}}

\bigskip
\bigskip             
In recent years, noncompact nuclear geometries of bubbles, tori, and
sheets have attracted considerable
interest\mycite{wong,bauer,moretto1,moretto2,borderie,batko,xu,moretto3}
in the context of nuclear multifragmentation studies.
According to the scenarios
considered,\mycite{bauer,moretto1,moretto2,borderie,batko,xu,moretto3}
it has been suggested that nuclear systems may assume transiently
exotic shapes, and then undergo 
a characteristic multifragment decay. One of the prominent
cases of noncompact geometries is that of an infinite sheet. This
case affords a high degree of computational simplicity in theoretical
modeling attempts but shows enough features common to many
exotic geometries to serve as a test ground for the validity of
various concepts.
As an example, it has been claimed\mycite{moretto1,moretto2}
that sufficiently thin sheets of nuclear matter, formed
dynamically during a heavy-ion collision, are subject to a new form of
instability driven by the proximity interaction of the
opposing surfaces. More recently, the concept of this sheet
instability has been applied\mycite{moretto3} to assess the stability of
Coulomb bubbles in general and against a {\it crispation mode}, 
in particular. In both these latter cases,
\mycite{moretto1,moretto2,moretto3}, the analysis relies critically 
on the liquid-drop model (LDM).
However, the assumptions of the LDM and, in particular, 
the various proposed sets of model 
parameters\mycite{seyler,myers1,myers2,myers3} 
were shown to be valid only for regular nuclear geometries 
but not for exotic ones. Similarly, the BUU\mycite{BUU}
or Landau Vlasov method,\mycite{LV}
used\mycite{bauer,borderie,batko,xu} in most of the theoretical
discussions of noncompact geometries are not designed to
handle quantum effects resulting from strong spatial
constraints associated with such geometries.

Below, the importance of quantum effects for the properties of nuclear
matter in noncompact spatial distributions is demonstrated for the
case of an infinite sheet. The scope of this study is limited  to the
calculation of the volume energy of symmetric nuclear matter. It is
clear, however, that also the surface energy would be affected by
quantum effects of the type considered, and that  the surface energy
needs to be calculated accordingly, before  realistic model
predictions can be made for the geometries of interest. Also, for the
sake of simplicity, the Coulomb energy is disregarded in the present
study. Central to the approach used in this work is the notion
of {\it bulk} matter as opposed to {\it surface} matter.
The former is characterized by a spatial uniformity in the
controlling parameters, and most notably, in the nucleonic momentum
distribution, while the latter is characterized by a nucleonic momentum
distribution that is changing rapidly with spatial coordinate
perpendicular to the surface. Only {\it bulk} matter 
is the subject of the present study. 

The energy per nucleon of {\it bulk} matter
can be calculated using a formalism
similar to that suggested by Seyler and Blanchard\mycite{seyler}
and employed successfully in the development of the
droplet model.\mycite{myers1,myers2,myers3} 
This formalism, based on 
the Thomas-Fermi approximation, was modified here to account 
approximately for 
effects of a quantization of the nucleonic momentum 
component $p_z$ perpendicular to the x-y surface plane
of the sheet. This quantization is a necessary consequence of 
the spatial confinement of the nucleonic motion
by the geometry of a sheet of a finite thickness $d$. 
As discussed further below, it is necessary to distinguish
between the {\it model} thickness $d$, used here to
construct the momentum distribution of nucleons, and the 
physical {\it matter} thickness $d_m$ describing
the profile of the corresponding spatial distribution of nuclear matter. 
The former quantity $d$ represents the width of an idealized
(square) confining potential well with infinitely high walls 
and defines the finite elementary quantum of the perpendicular 
momentum component $p_z$

\begin{equation}
\qquad \Delta p_z={h\over 2d}\;.
\label{eq_deltapz}
\end{equation}

As a result of the above quantization of $p_z$, single-nucleon states
for {\it bulk} sheet matter populate in nucleonic momentum space,
discrete, infinitely thin sheets at discrete values  of the
perpendicular momentum component, $p_z=k\Delta p_z$, where $k$ is any
nonzero integer. This type of population is in a clear contrast with
the uniform population of the Fermi sphere that is usually assumed in
Thomas-Fermi calculations\mycite{seyler,myers1,myers2,myers3}  modeling
nuclear matter. An example of such discrete population of the Fermi
sphere of a radius equal to the Fermi momentum $p_F$ is depicted in
Fig.~1. 

The quantization of the perpendicular component 
of the momentum can be be described by introducing a  population
function $f({\bf p},z)$,  defined as the density of nucleonic states
in momentum space per unit nuclear volume: 

\begin{equation}
\qquad f({\bf p},z)={4\Delta p_z\over h^3}
\Sigma_{k=1}^\infty\delta(p_z\pm \sqrt{k^2\Delta p_z^2-S(z)}\;)\;.
\label{eq_fp}
\end{equation}

\noindent
Here $\delta()$ denotes Dirac delta function, the factor
4 represents the spin-isospin degeneracy, and the 
function $S(z)$ describes 
the $z$-dependence of the momentum-independent part of the 
effective single-nucleon potential

\begin{equation}
\qquad S(z)=mod(U(z)-U(0),{\Delta p_z^2\over 2M})\;,
\label{eq_sz}
\end{equation}

\noindent
where $M$ denotes the average nucleon mass 
(taken as $M$=938.903 MeV/$c^2$). The origin of the
$z$ coordinate is set half way between the sheet surfaces. 
A boldface font is used in Eq.~\ref{eq_fp} and throughout this paper to 
denote a vector quantity. 

\vspace*{8.6cm}
\includegraphics{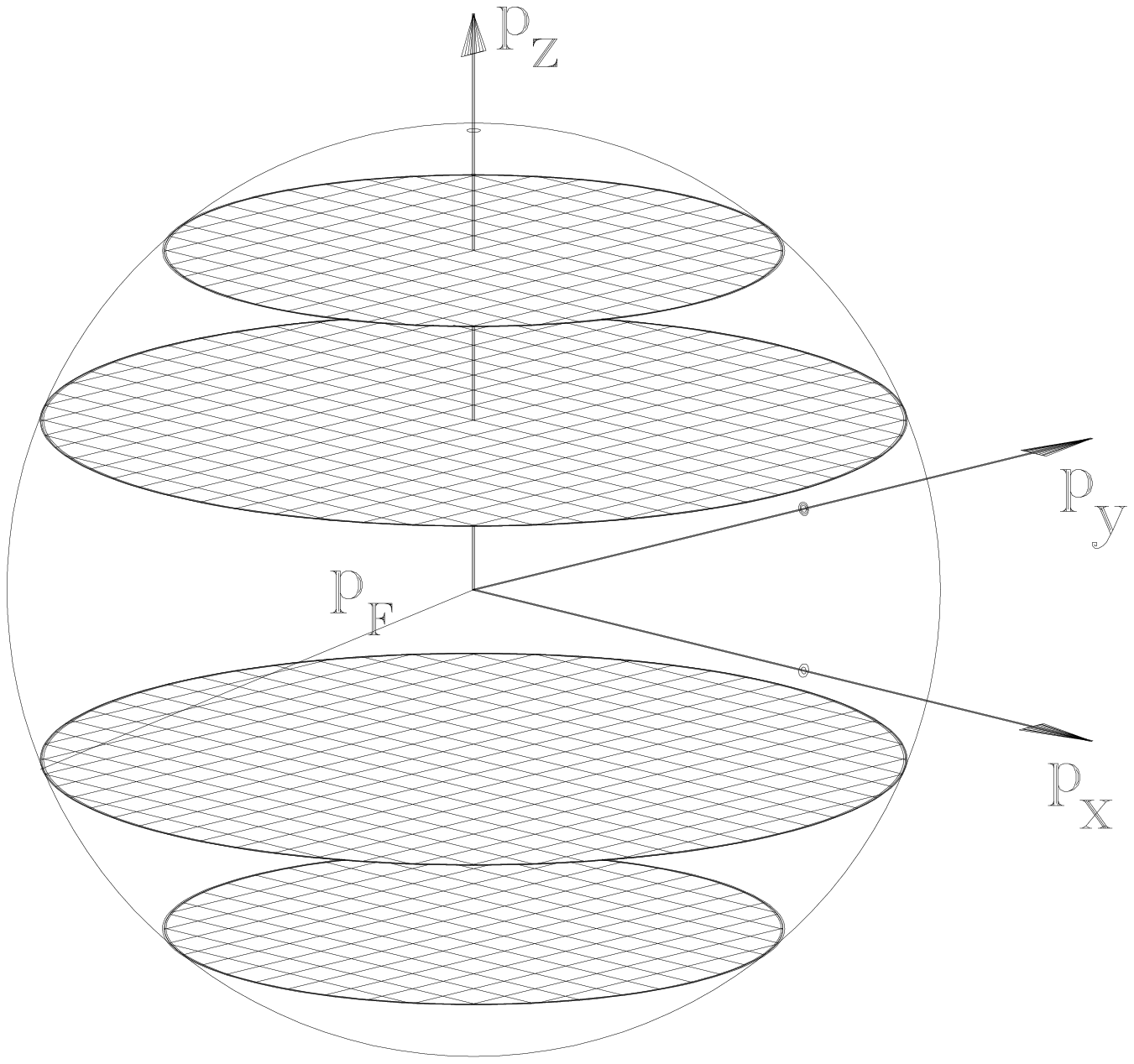}
{\footnotesize
\begin{quotation}
\noindent
Fig. 1. Population of nucleonic momentum space for {\it bulk}
matter of an idealized sheet, infinite in $x-y$ dimensions.
The z-axis is perpendicular to the surface of the sheet.
\end{quotation}}
\bigskip

For the nuclear matter density $\rho(z)$, the single-particle potential 
$U({\bf p},z)$, and the energy per nucleon 
$\epsilon_V(z)$, one writes then
in a close analogy to Refs.~\mycite{seyler,myers1,myers2,myers3}

\begin{equation}
\qquad \rho(z)=\int_{FS(z)}d{\bf p}f({\bf p},z)\;,
\label{eq_rhoint}
\end{equation}

\begin{equation}
\qquad U({\bf p},z)=\int d{\bf r'}\int_{FS(z)}d{\bf p'}f({\bf p'},z)
V(|{\bf r - r'}|,|{\bf p - p'}|)\;,
\label{eq_Up1}
\end{equation}

\noindent
and

\begin{equation}
\qquad \epsilon_V(z)={1\over \rho(z)}\int_{FS(z)}d{\bf p}f({\bf p},z)
[{p^2\over 2M}+{1\over 2}U({\bf p},z)]\;,
\label{eq_epsV}
\end{equation}

\noindent
respectively. In the above equations, $V(r,p)$ denotes the
nucleon-nucleon interaction, and $\int d{\bf r}$ and
$\int_{FS(z)}d{\bf p}$  denote integration over the (infinite) nuclear
volume and over the Fermi sphere of a radius $p_F(z)$ in momentum
space, respectively.  Assuming that the range of the nucleon-nucleon
interaction is small as compared to the linear dimensions of the {\it
bulk} domain, the spatial integration  in Eq.~\ref{eq_Up1} is
effectively limited to this domain. Then for the {\it bulk}
matter of interest here, all quantities in
Eqs.~\ref{eq_fp}--\ref{eq_epsV} are $z$-independent, in agreement
with the definition of the {\it bulk} matter.

The nucleon-nucleon interaction was assumed to be momentum dependent
and given by the Seyler-Blanchard\mycite{seyler} formula

\begin{equation}
\qquad V(r,p)=-C{e^{-r/a}\over (r/a)}(1-{p^2\over b^2})\;,
\label{eq_seyler}
\end{equation}

\noindent
where $C$ represents the strength of the interaction, 
the parameter $a$ represents
the range of the Yukawa force, and $b$ denotes a critical value of
the relative momentum, beyond which the force becomes repulsive. 
The quantities $r$ and $p$ are the distance between the nucleons
and their relative 
momentum, respectively. Values of the parameters of the interaction
were taken to be equal to\mycite{myers1} $C=328.61$~MeV, $a=0.62567$~fm, and
$b=392.48$~MeV/c. For infinite symmetric nuclear matter\mycite{myers1}, 
these values assure a volume energy per nucleon of $\epsilon_V$=-15.677 
MeV and a kinetic Fermi energy of 33.138 MeV.

A straightforward analytical integration
in Eqs.~\ref{eq_rhoint}, \ref{eq_Up1}, and \ref{eq_epsV},
using Eq~\ref{eq_seyler}, yields for the {\it bulk} matter
(the $z$ argument is left out for the sake of brevity)

\begin{equation}
\qquad \rho={8\pi\over h^3}(p_{zF}p_{F}^2-
{1\over 3}p_{zF}^3-{1\over 2}\Delta p_zp_{zF}^2)
-{1\over 6}(\Delta p_z)^2p_{zF}
\;,
\label{eq_rho}
\end{equation}

\begin{equation}
\qquad U(p)=V_o+V_1{p^2\over b^2}\;,
\label{eq_Up}
\end{equation}

\noindent
and

\begin{equation}
\qquad \epsilon_V=-{1\over 2}V_1+\epsilon_T{M\over M_{eff}}\;.
\label{eq_epsvol}
\end{equation}

\noindent
Here,

\begin{equation}
\qquad \epsilon_T={2\pi\over h^3\rho M}(p_{zF}p_{F}^4-
{1\over 5}p_{zF}^5-{1\over 2}\Delta p_z p_{zF}^4)
-{1\over 3}(\Delta p_z)^2p_{zF}^3+{1\over 30}(\Delta p_z^4)p_{zF}
\;
\label{eq_epsT}
\end{equation}

\noindent
is the average kinetic energy of a nucleon and

\begin{equation}
\qquad M_{eff}=M{b^2\over b^2+2M V_1}
\label{eq_meff}
\end{equation}

\noindent
is the effective mass summarizing effects of the interaction 
component quadratic in nucleonic momentum $p$. 
The strengths of the momentum-dependent and momentum-independent
components of the effective single-nucleon potential, respectively,
are given by

\begin{equation}
\qquad V_1=4\pi Ca^3\rho
\label{eq_V1}
\end{equation}

\noindent
and 

\begin{equation}
\qquad V_o=-V_1(1-{2M\over b^2}\epsilon_T)\;.
\label{eq_Vo}
\end{equation}

The quantity $p_{zF}$ in Eqs.~\ref{eq_rho} and \ref{eq_epsT} denotes
the maximum value of the perpendicular momentum $p_z$ allowed
for a given Fermi momentum $p_F$

\begin{equation}
\qquad p_{zF}=p_F-mod(p_F,\Delta p_z)\;.
\label{eq_pzf}
\end{equation}

Note that the corresponding equations for infinite nuclear matter,
characterized by a uniform population function  $f({\bf p},z)=4/h^3$,
can be readily obtained from the above equations \ref{eq_rho} --
\ref{eq_V1} by setting $p_{zF}$=$p_F$, while dropping all terms
containing powers of $\Delta p_z$.

The quantity of interest in the present study is the minimum value  of
the energy per nucleon, $\epsilon_V$, for {\it bulk} nuclear matter 
confined to the geometry of a sheet of finite thickness. 
This value can be obtained by
varying the input value of the Fermi momentum, $p_F$, in a search
routine minimizing $\epsilon_V$. Note that a calculation of $\epsilon_V$
for given $p_F$ and $d$ entails the use of
equations \ref{eq_deltapz},  \ref{eq_pzf}, \ref{eq_rho}, \ref{eq_V1},
\ref{eq_epsT}, \ref{eq_meff}, and \ref{eq_epsvol}
in an ordered sequence. Note also that in
the above equations, momenta are expressed in units of MeV/c, energies
in units of MeV, lengths in units of fm, and masses in units of
MeV/$c^2$. Accordingly, the Planck constant is in units of MeV fm/c,
$h$=1239.86 MeV fm/c.

Results of the calculations are summarized in Fig.~2, where values of
selected parameters characterizing properties of the {\it bulk} sheet
matter are plotted versus the {\it matter} thickness, $d_m$, of the
sheet. The latter quantity was evaluated for any given {\it model}
thickness $d$ based on the idealized matter density profile $\gamma
(z)$ along the direction perpendicular to the sheet surface.
The procedure of evaluating $d_m$ is illustrated in Fig.~3. The
density profile of interest was generated by (weighted) summing of the 
$sin^2[2\pi (z-d/2)p_z/h]$ functions for the actual distribution of
$p_z$.  The quantity $d_m$ was defined via the requirement that the 
density $\gamma (d_m/2)$ be equal to one-half of the bulk density
$\gamma_b$, where the latter was defined via the ``outermost'' (i.e.,
highest in $z<d/2$) solution of the equation

\begin{equation}
\qquad \int_0^Z\gamma (z)dz=Z\gamma (Z)=Z\gamma_b\;.
\label{eq_gammabulk}
\end{equation} 

The difference between the {\it model} and {\it matter} thicknesses
calculated in the above manner  depends on $d$. It increases linearly
from 0 to its maximum value of approximately 2.8 fm, with $d$
increasing from 0 to 5.6 fm. For $d>5.6$ fm, this difference decreases
quasi-hyperbolically with thickness, reaching
saturation at approximately 2 fm for $d>20 fm$. 

The main result of the present study is displayed in the top panel of 
Fig.~2. As seen in this panel, the maximum possible binding energy per
nucleon decreases dramatically as the thickness  $d_m$ of the sheet
decreases. This energy becomes negative, i.e., $\epsilon_V$ becomes
positive  and the system becomes unbound for thicknesses of about
$d_m\approx 2.8$ fm and below ($d<5.6$ fm). This effect of an increase 
in energy per nucleon with decreasing thickness of the sheet results
purely from the quantization of the perpendicular component $p_z$ of
the momentum (see Eq.~\ref{eq_fp}).  

\vspace*{16.3cm}
\includegraphics{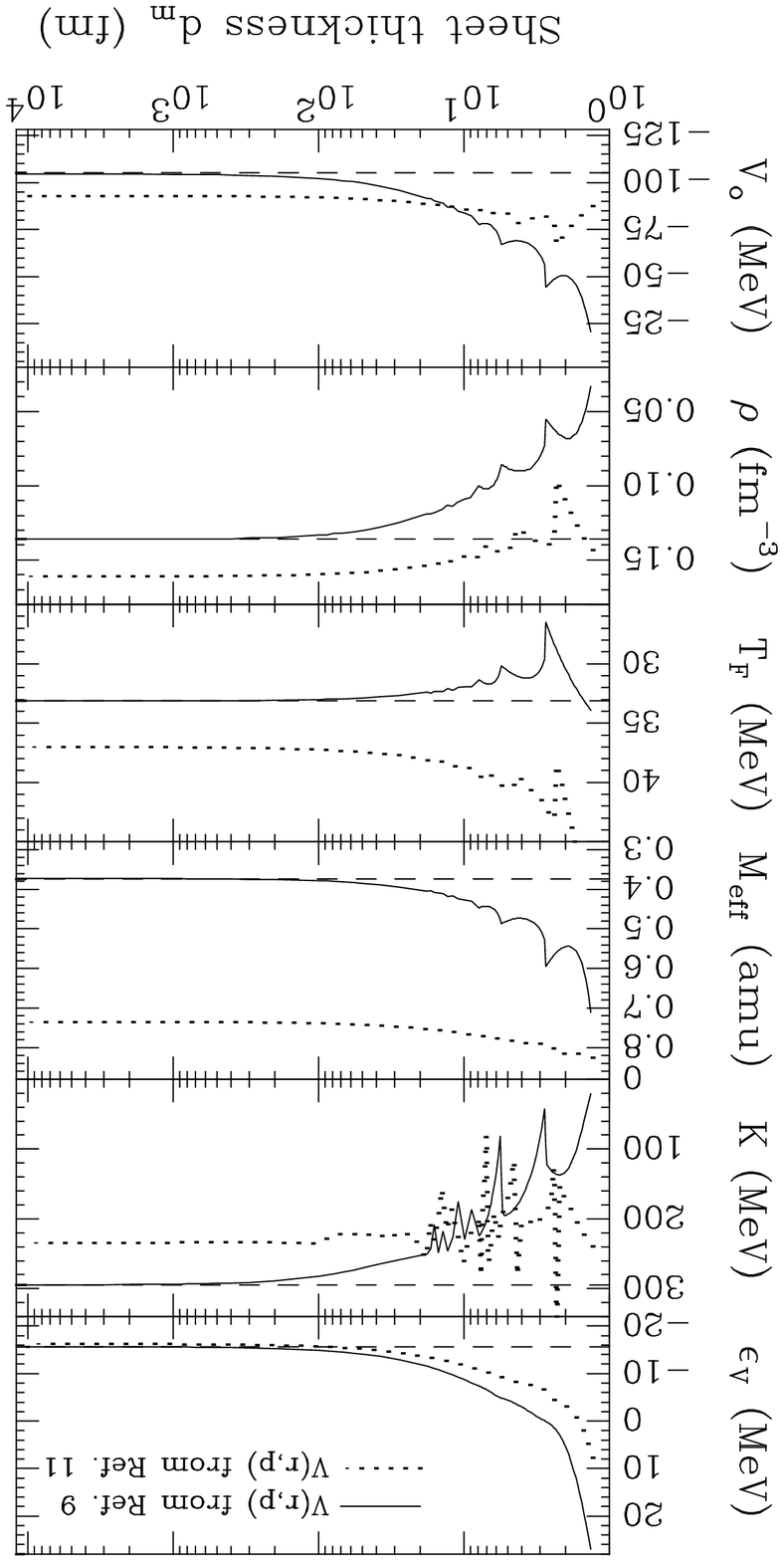} {\footnotesize \begin{quotation} 

\noindent 
Fig. 2. From top
to bottom: volume energy per nucleon, compressibility coefficient, 
reduced mass, nucleon kinetic Fermi energy, matter density,  
and strength of the momentum-independent single-particle potential, 
as functions of the sheet thickness $d_m$. The solid
curves are obtained with the Seyler-Blanchard$^{8}$
nucleon-nucleon interaction parameterization with values of
the parameters as indicated in the legend. Results obtained with the
parameterization of the interaction as proposed in Ref.~11 are shown
with dotted lines. 
\end{quotation}} 

\bigskip 

\noindent 
This feature is only approximately
accounted for by the standard droplet-model formula relying on a
uniform, Thomas-Fermi population of the Fermi sphere in momentum
space. The latter model ``adjusts'' the parameters of the
effective nucleon-nucleon interaction so as to
have the average shell effects included in the surface energy term.
The model relies here on the fact that
for compact shapes (e.g., of a square box)
the integrated (over the nuclear volume) shell effect is to a good
approximation proportional to the surface area. However, the large
magnitude of the volume shell effect raises the question
whether the
standard droplet model accounts accurately for
these effects in the casesof exotic shapes and whether it is, therefore,
applicable at all to nuclear matter in noncompact
geometries. The present paper does not answer fully this question.
At any rate, this {\it quantum} effect generates a strong
resistance of
nuclear matter against the development of noncompact geometries
associated with exotic spatial confinements of the nucleons motion,
in addition to that generated by a ``true'' surface tension.
It gives rise to strong effective forces,
akin to repulsive adiabatic
forces,\mycite{norenberg} driving the
nuclear system away from a formation of noncompact shapes attained
dynamically in nuclear reactions. 

\vspace*{8.0cm}
\includegraphics{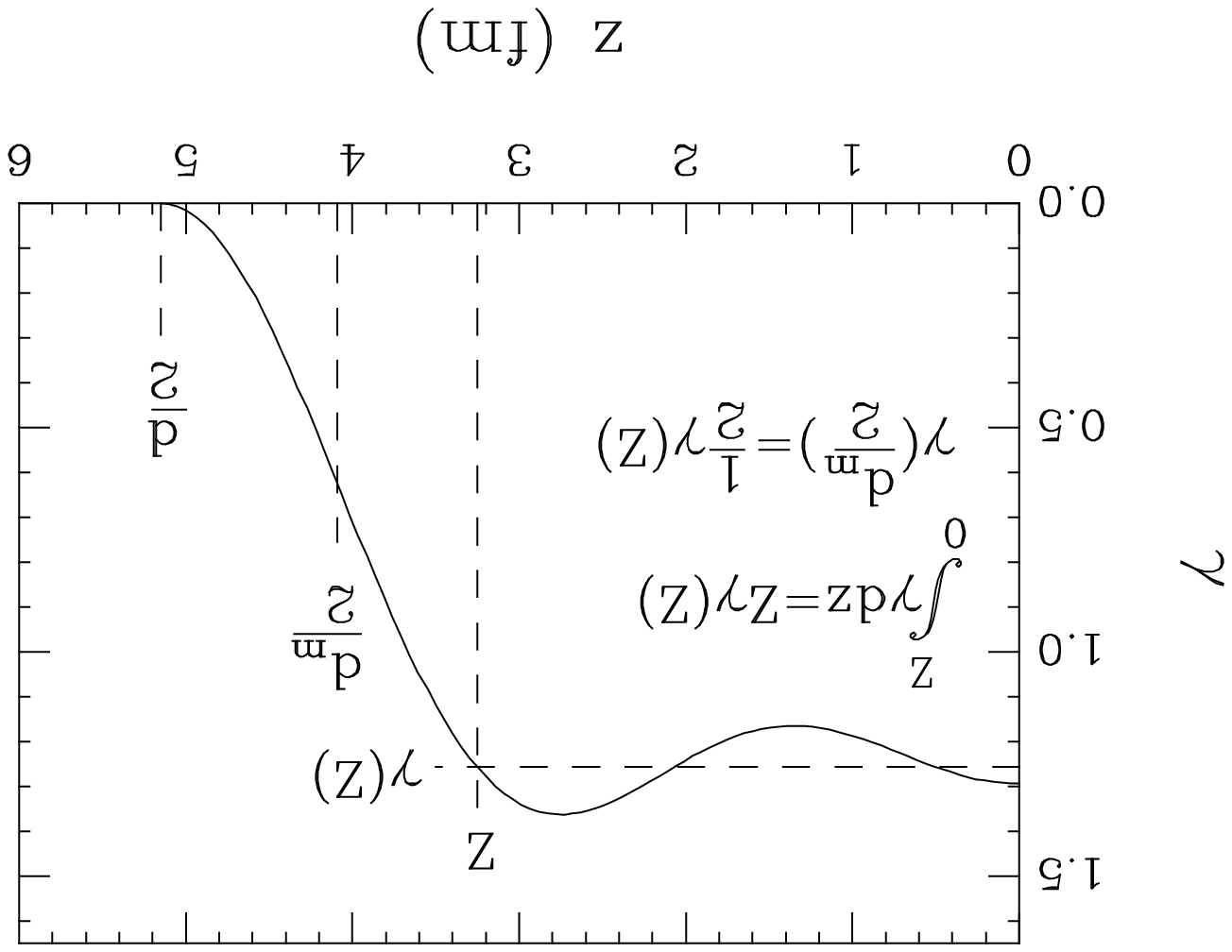}
{\footnotesize
\begin{quotation}
\noindent
Fig. 3. The relationship between the {\it model} and the {\it matter}
thickness of a sheet of nuclear matter.
\end{quotation}}
\bigskip

From the top panel of Fig.~2, one notes that the value of the
energy per nucleon for the {\it bulk} sheet matter
differs significantly from the asymptotic
value of $\epsilon_{V\infty}$=-15.7 MeV already for 
sheet thicknesses $d_m$ comparable in magnitude to nuclear diameters.
This indicates that the parameters of the Seyler-Blanchard
interaction\mycite{seyler,myers1} are not well suited
for absolute, more accurate calculations of the type reported
in the present work. The same is true with respect to any
parameterization that is set up so as to describe average trends
in nuclear masses and, hence, accounts for the average
shell effects due to a particular shape of a sphere, but not  
those associated with arbitrary shapes.

For comparison, in Fig.~2 
results obtained with a recently proposed\mycite{myers2,myers3}
parameterization of the nucleon-nucleon interaction
are shown as dotted lines. This new parameterization
adds two terms to the Seyler-Blanchard form --- an attractive
interaction inversely proportional to the relative momentum,
and a repulsive component proportional to the nuclear matter density
to the power of two-thirds.
The integrals involving the former
term were evaluated numerically. It is clear from the top panel
of Fig.~2 that the quantum effect discussed above depends on the
nucleon-nucleon interaction but that it  
is strong for either\mycite{myers1,myers2,myers3} parameterization.

The remaining four panels of Fig.~2,  shown mostly for the sake of
completeness, illustrate effects of a quantization of the
perpendicular momentum $p_z$ on selected properties of nuclear matter
at the stationary density minimizing the energy per nucleon of {\it
bulk} matter. Sharp dips and peaks in the respective functions result
from the  discontinuities in first derivatives of the underlying
momentum distribution at values of momenta $p$ that are integer
multiples of the the elementary quantum $\Delta p_z$. One notes, that
the new parameterization\mycite{myers2,myers3} results in asymptotic
values of the parameters that are different from those obtained using
the ``old''\mycite{myers1} parameterization, in agreement with those
reported in Ref.~\mycite{myers3} In particular, the values of
effective masses obtained with the new parameterization appear much
more realistic. 

In conclusion, the present analysis demonstrates the importance of
quantum effects for noncompact nuclear shapes. These effects are shown
to generate resistance of the nuclear systems against the development
noncompact  geometries and to generate forces driving these systems
toward compact shapes. The large magnitude of the effects discussed
above allows one to question the validity of the approximation
made in the standard liquied drop model, in which the average
(volume) shell effects of the type considered here are implicitly
included in the surface energy term. It allows one also to question
the applicability of BUU type of computations in cases where
noncompact geometries are involved. One notes in this latter respect
that, similarly to the Thomas-Fermi method discussed in this study,
the BUU equations do not consider effects of spatial confinement on
the spectrum of allowed states in momentum space. While it is not
clear, to what extent semiclassical models such as the modified
Thomas-Fermi approach followed in the present work, can capture the
essential features of strongly quantized systems,  more accurate
calculations, including surface and Coulomb energies, as well as
effects of a finite nuclear temperature seem desirable. A thorough
investigation  of the implications of the findings made in the present
work for several issues, appear clearly warranted. Problems of high
relevance to current reaction studies include question as to the
validity of BUU calculations, as well as issues associated with the
magnitude of realistic droplet-model parameters for nuclear matter in
general and not only for the specific case of spherical geometries of
real nuclei. For example, of relevance  in this context are the issues
concerning the relative stability of  the neck  matter between the
interacting nuclei and the decay modes of this matter. This problem is 
especially interesting for an understanding of the dynamical
production of intermediate-mass fragments\mycite{toke} in heavy-ion
reactions.

Illuminating discussions with Dr. W.J. Swiatecki are gratefully acknowledged.
This work was supported by the U.S. Department of Energy grant No.
DE-FG02-88ER40414.


\bigskip
\bigskip 
\begin{thebibliography}{99}

\bibitem{wong} C.Y. Wong, Ann. Phys. (N.Y.) {\bf 77}, 279 (1973). 

\bibitem{bauer} W. Bauer, G.F. Bertsch, and H. Schultz,
Phys. Rev. Lett. {\bf 69}, 1888 (1992).

\bibitem{moretto1}
L.G. Moretto, Kin Tso, N. Colonna, and G.J. Wozniak,
Phys. Rev. Lett. {\bf 69}, 1884 (1992).

\bibitem{moretto2} L.G. Moretto and G.J. Wozniak, Annu. Rev. Nucl. Part.
Sci. {\bf 43} 379 (1993), and references therein.

\bibitem{borderie} B. Borderie, B. Remaud, M.F. Rivet, and
F. Sebille, Phys. Lett. B {\bf 302}, 15 (1993); ibid B {\bf 307}, 404E
(1993).

\bibitem{batko} G. Batko and J. Randrup, Nucl. Phys. A {\bf 563},
97 (1993).

\bibitem{xu} H.M. Xu C.A. Gagliardi, R.E. Tribble, and C.Y. Wong,
Phys. Rev. C {\bf 49}, 1778 (1994). 

\bibitem{moretto3} L.G. Moretto, K. Tso, and G.J. Wozniak, 
Phys. Rev. Lett {\bf 78}, 824 (1997).

\bibitem{seyler} R.G. Seyler and C.H. Blanchard, Phys. Rev. {\bf 124}, 227
(1961); {\bf 131}, 355 (1963).

\bibitem{myers1} W.D. Myers and W.J. Swiatecki, Ann. Phys. {\bf 55}, 395
(1969).

\bibitem{myers2} W.D. Myers and W.J. Swiatecki, Ann. of Phys. {\bf
204}, 401 (1990); {\it ibid}, {\bf 211}, 292 (1991).                                              

\bibitem{myers3} W.D. Myers and W.J. Swiatecki, Nucl. Phys.
A {\bf 601}, 141 (1996).

\bibitem{BUU} G.F. Bertsch, H. Kruse, and S. Das Gupta, Phys. Rev. C
{\bf 29}, 673 (1984); G.F. Bertsch and S. Das Gupta, Phys. Rep. {\bf
160}, 189 (1988).

\bibitem{LV} C. Gregoire, et al., Nucl. Phys. A {\bf 465}, 317 (1987).

\bibitem{norenberg} W. N\"orenberg, Phys. Lett. B {\bf 104}, 107 (1981).

\bibitem{toke} J. T\~oke, et al., Phys. Rev. Lett. {\bf 77}, 3514 (1996).

\end{thebibliography}
\end{document}